\documentclass[preprint,nofootinbib,floats,eqsecnum,showpacs,11pt]{revtex4-1}
\usepackage{bm}
\usepackage{amsmath}
\usepackage{hyperref} 
\usepackage{color} 

\newcommand{\secref}[1]{Sec.~\ref{#1}}

\newcommand{\eqnref}[1]{(\ref{#1})}

\newcommand{\ket}[1]{\vert{#1}\rangle} 
\newcommand{\bra}[1]{\langle{#1}\vert} 

\begin{document}

\title{Larmor precession of the octet baryons in view of the general QCD parametrization}

\author{Dah-Wei Chiou}
\email{dwchiou@gmail.com}
\affiliation{Department of Physics and Center for Condensed Matter Sciences, National Taiwan University, Taipei 10617, Taiwan}

\author{Chun-Yen Lin}
\email{chunyenlin@cc.ncu.edu.tw} \affiliation{Department of Physics, National Central University, Jhongli, Taoyuan 32001, Taiwan}

\begin{abstract}
\textbf{We apply general QCD parametrization to describe an octet baryon at rest in a static and uniform magnetic field, in order to demonstrate a dynamical implementation of the hadron model. The derived evolution has the asymptote of Larmor precession, governed by an effective magnetic dipole moment coupling. The parameters appear in the effective magnetic dipole moment has been studied in the original works in kinematic settings. Here we show that the parameters have actual dynamical meaning, and are indeed measured by the process.
}
\end{abstract}

\pacs{12.38.Aw, 12.39.-x , 21.10.Ky , 23.20.-g }

\maketitle

\tableofcontents
\newpage

\section{Introduction}

With the success of the standard model, it is generally accepted that hadrons are bound states of quarks and gluons described by the relativistic field theory called quantum chromodynamics (QCD), but limited understanding about the nonperturbative physics of QCD has hindered first-principle calculations on the bound states.

 Nevertheless, many physical quantities of hadrons can be computed rather accurately by a simple nonrelativistic quark model (NRQM) in which a hadron (meson or baryon, resp.) is treated as a composite nonrelativistic quantum system composed of (2 or 3, resp.) quarks as spin-$1/2$ constituent particles. (NRQM is called the ``static quark model'' and detailed in $\S$2.7 of \cite{Wong:1998ex}.) The notable example is that the measured magnetic moments of the baryon octet can be fitted very precisely by only 2 or 3 parameters accounting for  the gyromagnetic ratios of constituent $u$, $d$, and $s$ quarks (in the 2-parameter fitting scheme, the gyromagnetic ratio of $d$ is identiﬁed as $−1/2$ of that of $u$) (see $\S$2.8 of \cite{Wong:1998ex} for details and \cite{Morpurgo:1989} for the 2-parameter scheme). On the other hand, dubbed the ``spin crisis'', polarized scattering experiments revealed that constituent quarks contribute surprisingly little to the proton's spin, which is in sharp conflict with the description of NRQM (see \cite{Jaffe:1995} and references therein).

It is conceivable that the underlying QCD could departure significantly from NRQM. The mystery is rather the other way around: Why does NRQM work so well for some hadron quantities, even though its foundational idea is refuted by polarized scattering experiments? Derived exactly from QCD under certain assumptions, the framework of ``general QCD parametrization'' (GP) is developed in \cite{Morpurgo:1989,Morpurgo:1990,Dillon:2009pf} to understand and compute many properties of hadrons. In terms of the hierarchy of parameters, GP is used to explain why NRQM is so successful and to predict when and how the results of NRQM are deviated from those of QCD.

The fundation of GP is the existence of a correspondence operator $V$ which maps every  NRQM hadron model state into its corresponding QCD hadron state.  $V$ is constructed as an adiabatic process of dressing up the model states with full QCD corrections, and it subsequently represents hadron state elements of QFT observables, with the corresponding elements of model operators. The nature of $V$ and the simplicity of the model space then lead to a finite and heirarchical parametrization of the exact hadron observables in terms of the NRQM operators, wherein the parameters are determined by the exact bound states. Existing literature \cite{Morpurgo:1989,Morpurgo:1990,Dillon:2009pf} had proved the kinematic correspondence powerful in accounting for the experimental data on hadron properties.

However, relatively few works are devoted to the dynamical aspects of GP. In addition to the correspondence between the exact and model operators, the relation between the operators and the measurable quantities is also crucial for the model. The latter correspondence can only be established in the context of the dynamical processes of measurements. Analysis on the dynamics can anchor the physical meaning of the model parameters, thereby enable GP to make detailed and testable predictions. To illustrate this point, we derive Larmor precession of the octet baryons by using general QCD parametrization in this paper.

Our goal is to establish a scheme to extract physical properties of hadrons from considering proper dynamical processes, with the application of GP. We will extract the dynamically defined magnetic dipole moments of the octet baryons, from their Larmor precessions caused by a static and uniform magnetic field. The matrix elements of the magnetic dipole moment operator were treated in \cite{Morpurgo:1989} based on a kinematic definition. Here we show that only the block-diagonal elements, each of which corresponds to a single baryon species, contribute to the Larmor precessions of the baryons. As a result, the dynamically defined magnetic dipole moment is parallel to the spin of the baryons, and is parameterized by the corresponding parameters discussed in \cite{Morpurgo:1989}.

\section{General QCD parametrization (GP)}
 In this section, we give a brief review on the GP formulation detailed in \cite{Dillon:2009pf}.
\subsection{General scheme of GP}\label{sec:general GP scheme}
Let $H_\mathrm{QCD}$ be the exact Hamiltonian of QCD, which contains the mass term for the $u$, $d$, $s$ quarks
\begin{equation}
\int d^3\mathbf{r}\left( m\left[\bar{u}_R(x)u_R(x)+\bar{d}_R(x)d_R(x)\right]
+(m+\Delta m)\bar{s}_R(x)s_R(x)\right),
\end{equation}
where the subscript $R$ refers to the mass renormalization point and $\Delta m$ is included for flavor breaking. Let $\ket{\Psi_B}$ be the eigenstate of $H_\mathrm{QCD}$ corresponding to the baryon $B$ in the rest system.\footnote{We review on the GP framework for baryons. It is modified in the obvious manner for mesons.} We have $H_\mathrm{QCD}\ket{\Psi_B}=M_B\ket{\Psi_B}$ and
\begin{equation}\label{MB}
M_B=\bra{\Psi_B} H_\mathrm{QCD} \ket{\Psi_B},
\end{equation}
with $M_B$ being the mass of the baryon $B$. The state $\ket{\Psi_B}$, being the exact eigenstate of a strongly interacting system of quarks and gluons, is a superposition of an infinite series of Fock space states. Schematically,
\begin{equation}
\ket{\Psi_B}=\ket{qqq}+\ket{qqqq\bar{q}}+\ket{qqq,\mathrm{gluons}}+\cdots,
\end{equation}
where the ellipsis stands for the infinite sum of additional states, and the coefficients for each state (depending on the momenta, spins, flavors, colors of intervening quarks, antiquarks and gluons) have been left out as unimportant.

Introduce the \emph{model Hamiltonian} $\mathcal{H}$ which acts only on the $3q$ sector. The only purpose of $\mathcal{H}$ is to provide a set of \emph{model states} $\ket{\Phi_B}$, which are to be mapped to the NRQM states $\Phi_B$ of 3 constituent quarks. In principle, we can relate the exact state $\ket{\Psi_B}$ with the model state $\ket{\Phi_B}$ as
\begin{equation}\label{Psi and Phi}
\ket{\Psi_B}=V\ket{\Phi_B},
\end{equation}
where $V$ is a very complicated unitary operator, which can be constructed in terms of $H_\mathrm{QCD}$ and $\mathcal{H}$ via the Gell-Mann and Low adiabatic construction of bound states (see Appendix I of \cite{Dillon:2009pf}). With $V$, the expression \eqnref{MB} becomes
\begin{equation}\label{MB 2}
M_B=\bra{\Phi_B} V^\dag H_\mathrm{QCD} V \ket{\Phi_B}.
\end{equation}

In \eqnref{MB}, the operator $H_\mathrm{QCD}$ is simple but the exact QCD state $\ket{\Psi_B}$ is complicated; in \eqnref{MB 2}, the model state $\ket{\Phi_B}$ is simple and all the complications are transformed into $V^\dag H_\mathrm{QCD}V$. Although $V^\dag H_\mathrm{QCD}V$ is indeed a very complicated operator, since it acts only on the degrees of freedom (space, spin, flavor, color) of the \emph{three} quarks in the model state $\ket{\Phi_B}$, it must be a function of these degrees only (after contraction of all creation and annihilation operators). Therefore, $V^\dag H_\mathrm{QCD}V$ behaves as a color singlet 3-body quantum mechanical operator acting on the three quarks of $\Phi_B$ numbered by 1, 2, 3.

The explicit form of $\Phi_B$ depends on how we select the model Hamiltonian $\mathcal{H}$. By choosing $\mathcal{H}$ to correspond to the simplest NRQM, the NRQM state $\Phi_B$ for the baryon of the $uds$ octet and decuplet takes the form in the nonrelativistic quantum mechanical description:
\begin{equation}\label{PhiB}
\Phi_B = X_{l=0}(\mathbf{r}_1,\mathbf{r}_2,\mathbf{r}_3)\, W_B(1,2,3)\, C(1,2,3),
\end{equation}
where $X_{l=0}(\mathbf{r}_1,\mathbf{r}_2,\mathbf{r}_3)$ is the space weve function of orbital angular momentum $\mathbf{L}=0$, $W_B(1,2,3)$ is the spin-flavor function which has the $SU(6)$ structure and total spin $S=1/2$ for the octet and $S=3/2$ for the decuplet, and $C(1,2,3)$ is the color singlet function. Correspondingly, in the relativistic field theory, the model state in the Fock space is given by
\begin{equation}\label{ket Phi}
\ket{\Phi_B} = \sum_{\mathbf{p},w} C^B_{w_1,w_2,w_3}(\mathbf{p}_1,\mathbf{p}_2,\mathbf{p}_3)\, a^\dag_{\mathbf{p}_1,w_1}a^\dag_{\mathbf{p}_2,w_2}a^\dag_{\mathbf{p}_2,w_2} \ket{0},
\end{equation}
where $\sum_{\mathbf{p},w}$ sums over all momenta $\mathbf{p}$'s and spin-flavor-color indices $w$'s, $a^\dag_{\mathbf{p}_i,w_i}$ are creation operators, and $C^B_{w_1,w_2,w_3}(\mathbf{p}_1,\mathbf{p}_2,\mathbf{p}_3)$, which contains a factor $\delta(\mathbf{p}_1+\mathbf{p}_2+\mathbf{p}_3)$, is obtained from $\Phi_B$ by performing the Fourier transform on $X_{l=0}(\mathbf{r}_1,\mathbf{r}_2,\mathbf{r}_3)$.
Intuitively, the transformation $V$ in \eqnref{Psi and Phi} can be understood as carrying out the tasks of
\begin{enumerate}
  \item dressing the three quark state $\ket{\Phi_B}$ with an infinite sum of of $q\bar{q}$ pairs and gluons;
  \item mixing the pure $l=0$ state of $\ket{\Phi_B}$ with $l\neq0$ states;
  \item transforming the Pauli 2-spinor states in $\Phi_B$ into the Dirac 4-spinor by completing the Pauli spinor with two zeros in the lower components.
\end{enumerate}

Because $\ket{\Phi_B}$ is a 3-quark state, the only part of $V^\dag H_{QCD}V$ that contributes in \eqnref{MB 2} is its projection on the $\ket{3q}$ sector of the Fock space:
\begin{equation}
\tilde{H}=\sum_{3q}\sum_{3q'} \ket{3q}\bra{3q} V^\dag H_{QCD}V \ket{3q'}\bra{3q'}.
\end{equation}
After contracting the creation and annihilation operators of $\tilde{H}$ with those in \eqnref{ket Phi}, the operator $\tilde{H}$ behaves as a quantum mechanical operator depending only on the space-spin-flavor variables of the three quarks. We call the 3-body quantum mechanical operator $\tilde{H}'$. Because both $H_\mathrm{QCD}$ and $\mathcal{H}$ transform as scalars under spatial rotation, the same are true for $V$ and $V^\dag H_{QCD}V$ and consequently $\tilde{H}'$ is a \emph{scalar} quantum mechanical operator. The number of independent scalar operators in the spin-flavor ($\boldsymbol{\sigma}$-$f$) space of the three quarks is of course finite, and we refer to them as $Y_\nu(\boldsymbol{\sigma},f)$, where the index $\nu$ specifies the operator we refer to. The most general form of $\tilde{H}'$ then is given by
\begin{equation}
\tilde{H}'=\sum_\nu R_\nu(\mathbf{r},\mathbf{r}')Y_\nu(\boldsymbol{\sigma},f),
\end{equation}
where $R_\nu(\mathbf{r},\mathbf{r}')$ are operators in the spatial coordinates of the three quarks with $\mathbf{r}\equiv(\mathbf{r}_1,\mathbf{r}_2,\mathbf{r}_3)$ and $\mathbf{r}'\equiv(\mathbf{r}'_1,\mathbf{r}'_2,\mathbf{r}'_3)$.
Consequently, the mass $M_B$ given by \eqnref{MB 2} reads as
\begin{equation}\label{MB 3}
M_B = \sum_\nu k_\nu \bra{W_B} Y_\nu(\boldsymbol{\sigma},f) \ket{W_B}
=:\bra{W_B}\text{``parameterized mass''}\ket{W_B},
\end{equation}
where the coefficients $k_\mu$ are
\begin{equation}
k_\nu=\bra{X_{l=0}(\mathbf{r})} R_\nu(\mathbf{r},\mathbf{r}') \ket{X_{l=0}(\mathbf{r}')}.
\end{equation}

After integrating over the spatial coordinates to obtain $k_\mu$, the ``parameterized mass'' is a function only of the spin and flavor operators of the three quarks. If we neglect electromagnetic corrections on the baryon masses, we then have the GP baryon mass formula for thee octet and decuplet:
\begin{eqnarray}\label{para mass}
\text{``parameterized mass''} &=&
M_0 +B\sum_iP^s_i +C\sum_{i>k}\boldsymbol{\sigma}_i\cdot\boldsymbol{\sigma}_k +D\sum_{i>k}\boldsymbol{\sigma}_i\cdot\boldsymbol{\sigma}_k(P^s_i+P^s_k)\nonumber\\
&&\mbox{} +E\sum_{i,k\neq j}^{i>k}\boldsymbol{\sigma}_i\cdot\boldsymbol{\sigma}_kP^s_j +a\sum_{i<k}P^s_iP^s_k +b\sum_{i<k}\boldsymbol{\sigma}_i\cdot\boldsymbol{\sigma}_kP^s_iP^s_k\nonumber\\
&&\mbox{} +c\sum_{i,k\neq j}^{i>k}\boldsymbol{\sigma}_i\cdot\boldsymbol{\sigma}_kP^s_j(P^s_i+P^s_k) +dP^s_1P^s_2P^s_3,
\end{eqnarray}
where $P^s_i$ denotes the projection onto the $s$ flavor for the $i$-th constituent quark (i.e.\ $P^s_is_i=s_i$, $P^s_iu_i=P^s_id_i=0$) and $M_0,B,C,D,E,a,b,c,d$ are constant parameters.

Fitted with the experimental baryon masses, the hierarchy of the parameters is evident: Numerical magnitudes of the parameters decrease rather strongly with the ``increasing complexity'' (i.e.\ complexity of the summand) of the term in \eqnref{para mass}.
Each $P^s_i$ in the summand, which accounts for the flavor breaking, gives rise to a reduction factor $\simeq0.3$. Additionally, each pair of different indices in the same $\sum$ corresponds to a ``gluon exchange'' and yields a reduction factor $\simeq0.37$. Furthermore, terms involving trace $\mathrm{Tr}$ are absent in the mass formula but present in the magnetic moment formula in \eqnref{para magnetic moment}-\eqnref{G nu}. Each trace $\mathrm{Tr}$ corresponds to a ``closed quark loop'' and gives rise to a comparatively much smaller reduction factor. We refer readers to $\S$7 of \cite{Dillon:2009pf} for more details and discussions on the hierarchy of the parameters.

By the hierarchy, we can ignore $a,b,c,d$ and keep only $M_0,B,C,D,E$. This 5-parameter approximation yields the NRQM results: the Gell-Mann-Okubo formula for the octet masses and the two equal spacing formulas of Gell-Mann for the decuplet masses ($\Omega-\Sigma^*=\Xi^*-\Sigma^*=\Sigma^*-\Delta$).

On the other hand, to account for the difference between $m_u$ and $m_d$ (e.g. $p^+\neq n$ for the masses), we have to take into consideration the electromagnetic corrections on baryon masses. In addition to the terms in \eqnref{para mass}, we will have more terms involving both $P^s_i$ and $Q_i$, the latter of which gives the charge ($+\frac{2e}{3}$ or $-\frac{e}{3}$) of the $i$-th quark. See $\S$10 of \cite{Dillon:2009pf} for more details.

The GP formulation can be straightforwardly applied to the lowest nonets of pseudoscalar ($J^\pi=0^-$) and vector ($J^\pi=0^-$) mesons. See $\S$8 of \cite{Dillon:2009pf}.

\subsection{Magnetic dipole moments}
The magnetic moment operator in the rest frame of the baryon is given the standard formula:
\begin{equation}
\boldsymbol{\mathcal{M}} =\frac{1}{2}\int d^3\mathbf{r}\left(\mathbf{r}\times\mathbf{j}(\mathbf{r},t)\right),
\end{equation}
where $\mathbf{j}(\mathbf{r},t)$ is the spatial part of the electromagnetic current $j_\mu(x)$:
\begin{equation}\label{j mu}
j_\mu(x) := e\left[
\frac{2}{3}\bar{u}(x)\gamma_\mu u(x)
-\frac{1}{3}\bar{d}(x)\gamma_\mu d(x)
-\frac{1}{3}\bar{s}(x)\gamma_\mu s(x)
\right].
\end{equation}
Parallel to \eqnref{MB}, the magnetic dipole moment of the baryon $B$ is given by
\begin{equation}\label{mu B}
\boldsymbol{\mu}_B=\bra{\Psi_B}\boldsymbol{\mathcal{M}}\ket{\Psi_B}.
\end{equation}
Following the same logic in \secref{sec:general GP scheme} with $H_\mathrm{QCD}$ replaced by $\boldsymbol{\mathcal{M}}$, we have, parallel to \eqnref{MB 3},
\begin{equation}\label{parameter g}
\boldsymbol{\mu}_B = \sum_\nu g_\nu \bra{W_B} \mathbf{G}_\nu(\boldsymbol{\sigma},f) \ket{W_B}
=:\bra{W_B}\text{``parameterized magnetic moment''}\ket{W_B},
\end{equation}
where $g_\nu$ are constant coefficients.

By the argument of spatial symmetry that $\boldsymbol{\mathcal{M}}$ is a pseudovector under spatial rotation, the magnetic moments for the $J^\pi=\frac{1}{2}^+$ octet baryons are, parallel to \eqnref{para mass}, given by
\begin{equation}\label{para magnetic moment}
\boldsymbol{\mu}_B = \sum_{\nu=0}^7 g_\nu \bra{W_B} \mathbf{G}_\nu \ket{W_B}
\equiv \sum_{\nu=1}^7 \tilde{g}_\nu \bra{W_B} \mathbf{G}_\nu \ket{W_B},
\end{equation}
where the $\mathbf{G}_\nu$'s are given by
\begin{eqnarray}\label{G nu}
&&\mathbf{G}_0=\mathrm{Tr}[QP^s]\sum_i\boldsymbol{\sigma}_i,
\quad
\mathbf{G}_1=\sum_iQ_i\boldsymbol{\sigma}_i,
\quad
\mathbf{G}_2=\sum_iQ_iP_i^s\boldsymbol{\sigma}_i,\nonumber\\
&&\mathbf{G}_3=\sum_{i\neq k}Q_i\boldsymbol{\sigma}_k,
\quad
\mathbf{G}_4=\sum_{i\neq k}Q_iP_i^s\boldsymbol{\sigma}_k,
\quad
\mathbf{G}_5=\sum_{i\neq k}Q_kP_i^s\boldsymbol{\sigma}_i,\nonumber\\
&&\mathbf{G}_6=\sum_{i\neq k}Q_iP_k^s\boldsymbol{\sigma}_i,
\quad
\mathbf{G}_7=\sum_{i\neq j\neq k}Q_iP_j^s\boldsymbol{\sigma}_k,
\end{eqnarray}
and satisfy the relation
\begin{equation}
\mathbf{G}_0=-\frac{1}{3}\mathbf{G}_1+\frac{2}{3}\mathbf{G}_2-\frac{5}{6}\mathbf{G}_3
+\frac{5}{3}\mathbf{G}_4+\frac{1}{6}\mathbf{G}_5+\frac{1}{6}\mathbf{G}_6
+\frac{2}{3}\mathbf{G}_7.
\end{equation}
The magnetic moments of the $J^\pi=\frac{1}{2}^+$ octet are specified by the parameters $\tilde{g}_1,\cdots,\tilde{g}_7$ as explicitly given in Equation (32) of \cite{Dillon:2009pf}.
Again, because of the hierarchy mentioned in \secref{sec:general GP scheme}, $\tilde{g}_1$ and $\tilde{g}_2$ are much larger than the remaining parameters $\tilde{g}_3,\cdots,\tilde{g}_7$. If we keep only $\tilde{g}_1$ and $\tilde{g}_2$ and ignore $\tilde{g}_3,\cdots,\tilde{g}_7$, \eqnref{para magnetic moment} reproduces the NRQM prediction with 2 parameters via the identifications:
\begin{equation}
\tilde{g}_1=-3\gamma_d\,\frac{\hbar}{2},
\qquad
\tilde{g}_2=3\left(\gamma_d-\gamma_s\right)\frac{\hbar}{2},
\end{equation}
where $\gamma_d=-\gamma_u/2$ and $\gamma_s$ are the 2 parameters used in NRQM with $\gamma_u,\gamma_d,\gamma_s$ accounting for the gyromagnetic ratios for $u$, $d$ and $s$ constituent quarks. 

Supposedly, the aforementioned GP formulation for magnetic moments can be carried over to the lowest decuplet ($J^\pi=\frac{3}{2}^+$) of baryons as well as to the lowest nonets of vector ($J^\pi=0^-$) mesons, although detailed investigation has yet to be done.

\subsection{Magnetic dipole ($M1$) transition of radiative decay}
NRQM works very well not only for hadron masses and magnetic moments but also for their radiative decays of the the magnetic dipole ($M1$) transition. For the decays from vector mesons into pseudoscalar mesons, see \cite{Morpurgo:1990}, $\S$9 of \cite{Dillon:2009pf}, and references therein; for the $\Sigma^0\rightarrow \Lambda^0\gamma$ decay, see $\S$VIII of \cite{Morpurgo:1989}; and for the $\Delta^+\rightarrow p^+\gamma$ dacay, see $\S$IX of \cite{Morpurgo:1989} and $\S$5 of \cite{Dillon:2009pf}. The GP method for radiative decays and the corresponding hierarchy of parameters are detailed in (particularly, $\S$9 of) \cite{Dillon:2009pf}. Here, we adopt the ideas of \cite{Dillon:2009pf} but, instead of deriving the GP parameters, investigate the relation between $M1$ transition rates and magnetic moments by following the treatment in $\S$2.4 of \cite{Sakurai}.

The QCD Lagrangian is given by
\begin{equation}
\mathcal{L}=\bar{q}\left(i\gamma^\mu D_\mu-m_q\right)q-\frac{1}{4}\mathbf{F}^{\mu\nu}\mathbf{F}_{\mu\nu},
\end{equation}
where $D_\mu$ is the covariant derivative with respect to the $SU(3)$ gauge field and $\mathbf{F}$ is the $SU(3)$ field strength. If we further consider the coupling to the $U(1)$ electromagnetic field, the Lagrangian is augmented as
\begin{equation}
\mathcal{L}=\cdots+\bar{q}\,\gamma^\mu\left(-q_e A_\mu\right)q-\frac{1}{4}{F}^{\mu\nu}{F}_{\mu\nu},
\end{equation}
where $A_\mu$ is the $U(1)$ gauge filed, $F$ is the $U(1)$ field strength, and $q_e$ is the quark's electromagnetic charge.
Consequently, the interaction Hamiltonian between the electromagnetic currents inside the baryon and the radiation field is given by (in $A^0=0$ gauge)
\begin{equation}\label{Hint}
H_\mathrm{int}=-\int d^3\mathbf{r}\
\mathbf{j}(\mathbf{r},t)\cdot\mathbf{A}(\mathbf{r},t)
\end{equation}
where $\mathbf{j}(x)$ is the spatial part of $j_\mu(x)$ defined in \eqnref{j mu}, $\mathbf{A}(x)$ is the quantized radiation field
given by
\begin{equation}
\mathbf{A}(\mathbf{r},t)=\frac{1}{\sqrt{2\pi}} \int d^3\mathbf{k}\sum_\alpha
\sqrt{\frac{\hbar}{2\,\omega}}
\left[
a_{\mathbf{k},\alpha}\mathbf{\epsilon}^{(\alpha)}e^{i\mathbf{k}\cdot\mathbf{r}-i\omega t}
+a^\dag_{\mathbf{k},\alpha}\mathbf{\epsilon}^{(\alpha)}e^{-i\mathbf{k}\cdot\mathbf{r}+i\omega t}
\right].
\end{equation}

For the emission of a photon characterized by the momentum $\mathbf{k}$ and polarization $\boldsymbol{\epsilon}^{(\alpha)}$ from the baryon state $\ket{\Psi_A}$ into $\ket{\Psi_B}$, we have
\begin{eqnarray}\label{A Hint B}
\bra{\Psi_B;n_{\mathbf{k},\alpha}+1}H_\mathrm{int}\ket{\Psi_A;n_{\mathbf{k},\alpha}}
=-\sqrt{\frac{(n_{\mathbf{k},\alpha}-1)\hbar}{4\pi\omega}}
\int d^3\mathbf{r}\,
\bra{\Psi_B}\, e^{-i\mathbf{k}\cdot\mathbf{r}}\,\mathbf{j}(\mathbf{r})\cdot\boldsymbol{\epsilon}^{(\alpha)}
\ket{\Psi_A}\, e^{i\omega t}.
\end{eqnarray}
In the framework of quantum field theory, \emph{spontaneous} emission ($n_{\mathbf{k},\alpha}=0$) and \emph{induced} (or stimulated) emission ($n_{\mathbf{k},\alpha}\neq0$) are treated in the same footing.

In the classical theory, as opposed to the quantum field-theoretic treatment, $\mathbf{A}$ is an externally applied potential which influences the charged currents but is not influenced by them. This classical description breaks down for the cases of small $n_{\mathbf{k},\alpha}$ and spontaneous emission. However, whenever the applied radiation field is intense enough, the classical description becomes satisfactory even within the framework of quantum field theory, since the occupation number $n_{\mathbf{k},\alpha}$ is so large that the radiation field can be regarded as an inexhaustible source/sink of photons. The equivalent classical vector potential used for an emission process is given by
\begin{equation}\label{A cl}
\mathbf{A}^{(\mathrm{class})}(\mathbf{r},t)
=\sqrt{\frac{(n_{\mathbf{k},\alpha}+1)\hbar}{4\pi\omega}}\
\boldsymbol{\epsilon}^{(\alpha)}\,e^{-i\mathbf{k}\cdot\mathbf{r}+i\omega t}
\approx \sqrt{\frac{n_{\mathbf{k},\alpha}\hbar}{4\pi\omega}}\
\boldsymbol{\epsilon}^{(\alpha)}\,e^{-i\mathbf{k}\cdot\mathbf{r}+i\omega t}
\end{equation}
for $n_{\mathbf{k},\alpha}\gg1$.

In typical electromagnetic decays of baryons, the wavelength of the emitted photon is much greater than the linear dimension of the baryon. This means we can replace $\exp(-i\mathbf{k}\cdot\mathbf{r})$ by the leading terms in the series:
\begin{equation}\label{exp ikr}
e^{i\mathbf{k}\cdot\mathbf{r}}
=1-i\mathbf{k}\cdot\mathbf{r} -\frac{(\mathbf{k}\cdot\mathbf{r})^2}{2} +\cdots.
\end{equation}
The first term in \eqnref{exp ikr} give rise to the \emph{electric dipole} ($E1$) transition (see \cite{Sakurai} for details). The matrix element \eqnref{A Hint B} involving the second term of \eqnref{exp ikr} is decomposed as
\begin{eqnarray}\label{E2+M1}
\bra{\Psi_B}(\mathbf{k}\cdot\mathbf{r})(\boldsymbol{\epsilon}^{(\alpha)}\cdot\mathbf{j})\ket{\Psi_A}
&=&\frac{1}{2} \bra{\Psi_B}(\mathbf{k}\cdot\mathbf{r})(\boldsymbol{\epsilon}^{(\alpha)}\cdot\mathbf{j}) +(\mathbf{k}\cdot\mathbf{j})(\boldsymbol{\epsilon}^{(\alpha)}\cdot\mathbf{r})\ket{\Psi_A}\nonumber\\
&&\mbox{}
+\frac{1}{2} \bra{\Psi_B}(\mathbf{k}\cdot\mathbf{r})(\boldsymbol{\epsilon}^{(\alpha)}\cdot\mathbf{j}) -(\mathbf{k}\cdot\mathbf{j})(\boldsymbol{\epsilon}^{(\alpha)}\cdot\mathbf{r})\ket{\Psi_A}.
\end{eqnarray}
The radiative transition due to the first term of \eqnref{E2+M1}, which can be written as
\begin{equation}
(\mathbf{k}\cdot\mathbf{r})(\boldsymbol{\epsilon}^{(\alpha)}\cdot\mathbf{j}) +(\mathbf{k}\cdot\mathbf{j})(\boldsymbol{\epsilon}^{(\alpha)}\cdot\mathbf{r})
=\mathbf{k}\cdot\left(\mathbf{x}\,\mathbf{j}+\mathbf{j}\,\mathbf{x}\right)\cdot
\boldsymbol{\epsilon}^{(\alpha)},
\end{equation}
is known as the \emph{electric quadrupole} ($E2$) transition (see \cite{Sakurai} for details). On the other hand, the radiative transition due to the second term of \eqnref{E2+M1}, which can be written as
\begin{equation}
(\mathbf{k}\cdot\mathbf{r})(\boldsymbol{\epsilon}^{(\alpha)}\cdot\mathbf{j}) -(\mathbf{k}\cdot\mathbf{j})(\boldsymbol{\epsilon}^{(\alpha)}\cdot\mathbf{r})
=(\mathbf{k}\cdot\boldsymbol{\epsilon}^{(\alpha)})\cdot(\mathbf{r}\times\mathbf{j}),
\end{equation}
is known as the \emph{magnetic dipole} ($M1$) transition.

In the case of spontaneous emission $A\rightarrow B+\gamma$ (a hadron $A$ makes a radiative transition into a different hadron $B$), the transition probability per unit time into a solid angle element $d\Omega_\mathbf{k}$ is given by the famous \emph{Golden rule} (see \cite{Sakurai} for details):
\begin{eqnarray}\label{Golden rule}
w_{d\Omega} &=& \frac{2\pi}{\hbar}
\Big|\bra{\Psi_B;n_{\mathbf{k},\alpha}=1}H_\mathrm{int}\ket{\Psi_A;n_{\mathbf{k},\alpha}=0}\Big|^2
\frac{\omega^2}{(2\pi)^2}\frac{d\Omega_\mathbf{k}}{\hbar}\nonumber\\
&=&\frac{\omega}{8\pi^2\hbar}
\left|\int d^3\mathbf{r}\,\bra{\Psi_B}\,
e^{-i\mathbf{k}\cdot\mathbf{r}}\,\mathbf{j}(\mathbf{r})\cdot\boldsymbol{\epsilon}^{(\alpha)}
\ket{\Psi_A}
\right|^2 d\Omega_\mathbf{k},
\end{eqnarray}
where $\omega$ is the frequency of the emitted phone, which satisfies $M_A=M_B+\hbar\omega$.
The $M1$ transition rate is given by
\begin{eqnarray}
w_{d\Omega}^{(M1)} &=& \frac{\omega}{8\pi^2\hbar}
\left|\int d^3\mathbf{r}\,\bra{\Psi_B}\,
(\mathbf{k}\cdot\boldsymbol{\epsilon}^{(\alpha)})\cdot(\mathbf{r}\times\mathbf{j})
\ket{\Psi_A}
\right|^2 d\Omega_\mathbf{k}\nonumber\\
&\equiv&
\frac{\omega}{4\pi^2\hbar}
\left|(\mathbf{k}\cdot\boldsymbol{\epsilon}^{(\alpha)})\cdot
\bra{\Psi_B}\boldsymbol{\mathcal{M}}\ket{\Psi_A}
\right|^2d\Omega_\mathbf{k}.
\end{eqnarray}
In the same spirit of \eqnref{mu B}, the relevant matrix element $\bra{\Psi_B}\boldsymbol{\mathcal{M}}\ket{\Psi_A}$ can be regarded as the ``off-diagonal'' part of the magnetic dipole moment:
\begin{equation}
\boldsymbol{\mu}_{A\rightarrow B}=\bra{\Psi_B}\boldsymbol{\mathcal{M}}\ket{\Psi_A}
\equiv\bra{\Phi_B}V^\dag\boldsymbol{\mathcal{M}}V\ket{\Phi_A}
\end{equation}
In parallel to \eqnref{parameter g}, $\boldsymbol{\mu}_{A\rightarrow B}$ can be parameterized by the GP scheme:
\begin{equation}
\boldsymbol{\mu}_{A\rightarrow B}
=\bra{W_B}\text{``parameterized magnetic moment''}\ket{W_A}.
\end{equation}
This explains why NRQM works very well for the radiative decays of the $M1$ transition from the perspective of GP.

\section{Larmor precession and magnetic dipole moments}

Although we have understood why NRQM is successful for the ``diagonal'' part of the magnetic moment $\boldsymbol{\mu}_B$ given by \eqnref{mu B}, we do not know the exact \emph{dynamical} relevance of $\boldsymbol{\mu}_B$, in the sense that the off-diagonal magnetic moment $\boldsymbol{\mu}_{A\rightarrow B}$ corresponds to the $M1$ radiative decay. To understand the dynamical meaning of $\boldsymbol{\mu}_B$, we repeat the above analysis for the case of spin precession of a hadron subject to a classical electromagnetic field, in parallel to the case of spontaneous emission.

It is known that the octet baryons do not decay strongly. Ignoring the weak interaction, we may thus treat these baryons as ``free'' QCD systems subject to the electromagnetic perturbation. For this simplicity we will consider only the octet baryons in this paper, despite the applicability of GP on other types of hadrons. The “free Hamiltonian” operator in our case is ${H}_0\equiv {H}_\mathrm{{QCD}}$. The ``interaction Hamiltonian" is given by \eqnref{Hint}, but with the field $\mathbf A$ set to be a static classical field $\mathbf{A}^{{(\mathrm{class})}}(\mathbf r)$. Thus we have: 
\begin{equation}
{H}_\mathrm{int}=\int d^3\mathbf{r}\,\,\mathbf{A}^{(\mathrm{class})}(\mathbf r)\cdot {\mathbf{j}}(\mathbf{r}).
\end{equation}
The full Hamiltonian is $ H={H}_{0}+{H}_\mathrm{int}$ for our system of quarks and gluons. In this paper we will restrict to the case with a static and uniform magnetic field $\mathbf{B}(\mathbf{r},t)= \mathbf{B}_0$, so $\mathbf{A}^{(\mathrm{class})}(\mathbf{r})=\frac{1}{2} \mathbf{B}_0\times\mathbf{r}$. For this case we have:
\begin{equation}\label{Hint B}
{H}_\mathrm{int}=\mathbf{B}_0\cdot\int d^3\mathbf{r}\, \left(\mathbf{r}\times {\mathbf{j}}(\mathbf{r})\right).
\end{equation}

In order to focus on the spin degrees of freedom, we want to discuss the octet baryon states at rest.  Thus, for consistency we assume $ H_\mathrm{int}$ does not mix the initial octet baryon states at rest with the states having nonzero total momentum. This assumption is consistent with experimental facts, and it allows us to consider the evolution in the zero-momentum subspace of the full QCD Hilbert space, which is spanned by the basis :
\begin{equation}\label{basis}
\begin{split}
 {\mathbf{I}}_{(P=0)}= \sum_{i=1}^{\mathcal N} \ket{\Psi_{i} } \bra{\Psi_{i}}+\int_{\mathcal N +1}^{\infty} d\rho \ket{\Psi_{\rho} } \bra{\Psi_{\rho}}\equiv \sum^\mathrm{dist}_m \ket{\Psi_{m} } \bra{\Psi_{m}} \\
\end{split}
\end{equation}
where $\{\ket{\Psi_{i} }\}$ includes single-particle states and bound states, and $\{\ket{\Psi_{\rho}}\}$ includes multi-particle states, and we introduced the notation $\sum^\mathrm{dist}_m$ for the expansion with the index $m$ running over both $i$ and $\rho$. Also, we have $\hat{H}_{0} \ket{\Psi_{m} }= E_m \ket{\Psi_{m} }$. Note that we expect the energy seperation $E_{\rho}>  E_i $ between the discrete and continuous parts of the spectrum.
We also choose the spin-1/2 octet baryon states to be the first sixteen members of the basis, setting $\Psi_{2N-1}$ and $\Psi_{2N}$ to be the spin-up and spin-down states of the $Nth$ octet baryon ($N=1,\cdots,8$).

Let us now evaluate the transition amplitude from a spin-up baryon state $\Psi_{2N}$ to an arbitrary state $\Psi_{j}$ from the basis, assuming the perturbation $H_\mathrm{int}$ is switched on over a period of time $t$: 
\begin{equation}\label{trans amp 1}
\begin{split}
\bra{\Psi_j } {U}(0,t) \ket{\Psi_{2N}}
=\bra{\Psi_j}e^{(-i  t{H}_{0}/\hbar)} {U}_I(0,t) \ket{\Psi_{2N}}
=e^{-iE_jt/\hbar} \bra{\Psi_j} {U}_I(0,t) \ket{\Psi_{2N}}
\end{split}
\end{equation}
where
\begin{equation}
\begin{split}
{U}_I(0,t)\equiv \mathcal{T}\exp\left(-\frac{i}{\hbar}\int_0^t  H^I_\mathrm{int}(t') dt' \right)\,,\,\,
 H^I_\mathrm{int}(t)\equiv e^{(i  {H}_{0}t/\hbar)} H_\mathrm{int}\,e^{(-i {H}_{0}t/\hbar)}.
\nonumber
\end{split}
\end{equation}
We then expand ${U}_I(0,t)$ with Dyson series:
\begin{eqnarray}\label{trans amp 2}
\bra{\Psi_j } {U}(0,t) \ket{\Psi_{2N}}
&=&e^{-iE_jt/\hbar}\bra{\Psi_j }\sum_{K=0}^\infty \left(\frac{-i}{\hbar}\right)^K\prod_{n=1}^K \int_0^{t_{n-1}}dt_n \, H^I_\mathrm{int}(t_n) \ket{\Psi_{2N}}\nonumber\\
&\equiv& e^{-iE_jt/\hbar}\sum_{K=0}^\infty M_K
\end{eqnarray}
with $t_0\equiv t$. We may now sandwich each factor of $H^I_\mathrm{int}(t_n)$ in $M_K$ with the basis \eqnref{basis} and obtain:
\begin{eqnarray}\label{M_K}
M_K
&=&\left(\frac{-i}{\hbar}\right)^K \left(\sum_{m_0}\sum_{m_1}\cdot\cdot\cdot\sum_{m_{K}}\right) \langle{\Psi_j}|{\Psi_{m_0}}\rangle\langle\Psi_{m_K}|\Psi_{2N}\rangle\nonumber\\
&&\times\prod_{n=1}^K \int_0^{t_{n-1}}dt_n \, e^{-i  t_n(E_{m_n}-E_{m_{n-1}}) /\hbar}  \,\bra{\Psi_{m_{n-1}}} H_\mathrm{int} \ket{\Psi_{m_n}}
\end{eqnarray}
Next, we assume that baryon degeneracy in the spectrum of ${H}_\mathrm{QCD}$ exists only between the two spin states of a single baryon species , e.q. $E_{2N-1}=E_{2N}$. Accordingly, we then split the basis into a subbasis $\{\Psi_{\bar m}\}\equiv\{\Psi_{2N}, \Psi_{2N-1}\}$ for the $N$th baryon, and that for all other states $\{\Psi_{m'}\}|_{E_{ m'}\neq E_{2N}}$. Clearly, the contributions to $M_K$ involving only intermediate states from $\{\Psi_{\bar m}\}$ will grow as $t^K$ with time. On the other hand, the contributions with at least one of the intermediate states from $\{\Psi_{m'}\}$ will have at least one of the time integrals regularized by the exponential factors,  and only grow as  $t^{K-1}$ or slower. To separate the two kinds of contributions, we introduce 
\begin{equation}\label{N projection}
H_\mathrm{Lar}\equiv\sum_{N} {P}^{\dag}_N \,H_\mathrm{int}\, {P}_N
\end{equation}
with the projection operator $P_N$ defined as
\begin{equation}
{P}_N\equiv \ket{\Psi_{2N} } \bra{\Psi_{2N}}+\ket{\Psi_{2N-1} } \bra{\Psi_{2N-1}},
\end{equation}
and find
\begin{eqnarray}\label{M_K}
M_K= \frac{1}{K!}\left(\frac{-i\,t}{\hbar}\right)^K\bra{\Psi_j}(H_\mathrm{Lar})^K \ket{\Psi_{2N}}+O(  \tau_{{}_{(N)}}\,t^{K-1})
\end{eqnarray}
where in the second term a factor of $t$ is replaced with  $\tau_{{}_{(N)}}\equiv \max\{\hbar/|E_i-E_{2N}|\}_{i\neq 2N, 2N-1}$ as a result of the regularization. Note that when $t\gg\tau_{{}_{(N)}}$, the $O( \tau_{{}_{(N)}}\,t^{K-1})$ term becomes negligible compared with the first term in $M_K$, and this leads to the result:
\begin{eqnarray}\label{trans amp 3}
\lim_{{t}/{\tau_{{}_{(N)}}}\to \infty}\bra{\Psi_j } {U}(0,t) \ket{\Psi_{2N}}
&=&\lim_{{t}/{\tau_{{}_{(N)}}}\to \infty}e^{-iE_it/\hbar}\bra{\Psi_j } e^{-i H_\mathrm{Lar}t/\hbar} \ket{\Psi_{2N}}\nonumber\\
&=&\lim_{{t}/{\tau_{{}_{(N)}}}\to \infty}\bra{\Psi_j } e^{-i \left(H_0+H_\mathrm{Lar}\right)t/\hbar} \ket{\Psi_{2N}}\,,
\end{eqnarray}
where we have used $[H_0,H_\mathrm{Lar}]=0$. Since the final basis state $\Psi_j$ is arbitrary, we have obtained the asymptotic evolution of the state $\Psi (t)$ with $\Psi (0)=\Psi_{2N}$:
\begin{eqnarray}\label{evolution}
\lim_{{t}/{\tau_{{}_{(N)}}}\to \infty} \ket{\Psi (t)}
&=&\lim_{{t}/{\tau_{{}_{(N)}}}\to \infty} e^{-i \left(H_0+ H_\mathrm{Lar}\right)t/\hbar} \ket{\Psi(0)}\,,
\end{eqnarray}
or equivalently, for an arbitrary state $\Psi(0)$ in the subspace of octet baryons with zero momentum:
\begin{eqnarray}\label{shrodinger eq}
 \lim_{{t}/{\tau_{{}_{(N)}}}\to \infty}  \left(H_\mathrm{QCD}+H_\mathrm{Lar}\right)\ket{\Psi(t)}
&=& \lim_{{t}/{\tau_{{}_{(N)}}}\to \infty}  i\hbar\frac{\partial}{\partial t}\ket{\Psi(t)}.\nonumber\\
\end{eqnarray}
Note that $H_0+H_\mathrm{Lar}$ is the effective Hamiltonian governing the dynamics after $t\gg \tau_{{}_{(N)}}$, and
the time-dependent Schr\"{o}dinger equation describes a species-preserving evolution in the subspace of octet baryons. For the details of the evolution, we need further knowledge about $H_\mathrm{Lar}$.

Now we have come to the crucial step of our calculation. Since $\lim_{{t}/{\tau_{{}_{(N)}}}\to \infty} \ket{\Psi(t)}$ is a baryon state, we have $\lim_{{t}/{\tau_{{}_{(N)}}}\to \infty} \ket{\Psi(t)}= V\ket{\Phi(t)}$. This allows us to use GP, and the effective Hamiltonian operator $H_0+H_\mathrm{Lar}$ is given by the model operator $ V^{\dag} \left(H_\mathrm{QCD}+H_\mathrm{Lar}\right) V$ with finite set of hierarchical parameters. Then, \eqnref{shrodinger eq} implies:
\begin{eqnarray}
\left(V^{\dag} H_\mathrm{QCD} V+V^{\dag}  H_\mathrm{Lar} V\right)\ket{\Phi(t)}
&=&  i\hbar\frac{\partial}{\partial t} \ket{\Phi(t)}.
\end{eqnarray}
We may expand $\ket{\Phi(t)}$ with the two model spin stastes $\{\ket{\Phi_{I}}\equiv V^{\dag} \ket{\Psi_{I}}\}_{I=2N,2N-1}$ of the $N$th baryon:
\begin{equation}
\ket{\Phi(t)} =\sum_{I} e^{-iE_{I}t/\hbar} C_{I}(t) \ket{\Phi_{I}}.
\end{equation} 
One can see clearly now, the elements $\bra{\Phi_{I'}}V^{\dag} H_\mathrm{QCD} V\ket{\Phi_{I}}= \delta_{I'I}E_I$ are given by the parametrized mass $M_{B=N}$ elements defined in \eqnref{MB 3}. Also, from \eqnref{N projection} and \eqnref{Hint B} we imediately see that the elments $ \bra{\Phi_{I'}}V^{\dag}  H_\mathrm{Lar} V\ket{\Phi_{I}}= \dot{C}_{I'}(t)/ C_{I}(t) $ are given by the elements $ \mathbf{B}_0\cdot\boldsymbol{\mu}_{B=N}$ defined in \eqnref{mu B}.

The result implies that, as a prediction of GP, Larmor precession is the asymptotic behavior of an octet baryon $(N)$ at rest in the static and uniform magnetic field, after the evolution time satisfies $t\gg\tau_{{}_{(N)}}$. Therefore, the dynamical meaning of $\boldsymbol{\mu}_B$ is the magnetic dipole moment of the hadron $B$ measured by the Larmor precession. Moreover, according to \eqnref{para magnetic moment} and \eqnref{G nu}, the magnetic dipole moment can be parameterized by 7 parameters $\tilde{g}_\nu$ and, as a consequence, $\boldsymbol{\mu}_B$ is parallel to the baryon's spin.

\section{Conclusions}

We had reviewed the method of QCD general parametrization, which prescribes the kinematic correspondence between the exact QCD hadron states and the NRQM model states through the operator $V$. Consequently, the hadron-state matrix elements of a QFT operator are given by the elements of the corresponding NRQM operator, with a finite set of hierarchical parameters.

As shown by our analysis, this kinematic correspondence has dynamical implications. Particularly, it may predict the evolution of an octet-baryon in an external weak classical field. With the above derivation as an example, we propose the following general scheme.

The initial octet baryon state is evolved by $H_\mathrm{QCD}+H_\mathrm{int}$, where $H_\mathrm{int}$ results from the external field. The standard perturbation calculation gives the evolution, in terms of the transition amplitudes induced by $H_\mathrm{int}$ between the eigenstates of $H_\mathrm{QCD}$. From the Fermi golden rules, we expect the dominating terms to come from the transitions between the states close to degeneracy with the initial state. Therefore, if the initial state is a slowly moving octet baryon, the dominating terms will be given by the transition amplitudes between only the baryon states. These are the matrix elements of the effective Hamiltonian for the evolution, and they lie in the domain of GP . Under GP, the effective Hamiltonian corresponds to a model Hamiltonian in NRQM, and the evolution of the NRQM states gives the evolution of the exact states. Moreover, the finite set of parameters in the model Hamiltonian, determined by the inner structures of the baryons, acquire concrete dynamical meaning as being measured by the physical process.

In this paper we demonstrate the scheme by deriving the evolution from an initial octet baryon state at rest, in a static and uniform magnetic field. We had assumed the total momentum remains zero during the evolution. We find the effective Hamiltonian $H_\mathrm{Lar}$ to be the species-diagonal projection of $H_\mathrm{int}$. Through GP, we also find the evolution to be the Larmor precession under the magnetic field. Thus the parameters in the effective Hamiltonian, which have been studied in \cite{Morpurgo:1989}, have the meaning of giving the effective magnetic dipole-moment as measured by the Larmor precession.

Finally, we expect the scheme to be valid in a wide range of dynamical processes involving slowly moving baryons in weak external fields. It is our hope that further studies in this direction would bring more insight into the meanings and implications of the hierarchical parameters from GP, in the dynamical sense.

\begin{acknowledgments}
The authors are grateful to Professor Wei-Tou Ni for valuable advice and inspiring discussions. Dah-Wei Chiou was supported in part by the Ministry of Science and Technology of Taiwan under the Grant No.\ 101-2112-M-002-027-MY3; Chun-Yen Lin was supported in part by the Ministry of Science and Technology of Taiwan under the Grant No.\ 102-2112-M-008 -015 -MY3.
\end{acknowledgments}

\end{document}